\documentclass[NumberedRefs,Preprint,10pt]{JASA}
\usepackage[utf8]{inputenc}
\usepackage[english]{babel}
\usepackage{wrapfig}
\usepackage{amsmath}
\usepackage{amsfonts}
\usepackage{amssymb}
\usepackage{makeidx}
\usepackage{graphicx}
\usepackage{lmodern}
\usepackage{hyperref}
\usepackage{mathtools}
\usepackage{booktabs}
\usepackage{rotating}
\usepackage{cleveref}
\usepackage{setspace}
\usepackage[left=2cm,right=2cm,top=2cm,bottom=2cm]{geometry}
\usepackage{comment}
\usepackage{multirow}

\let\orgautoref\autoref

\renewcommand{\autoref}[1]
{%
\def\figureautorefname{Fig.}%
\def\subfigureautorefname{Fig.}%
\def\equationautorefname{Eq.}%
\orgautoref{#1}%
}

\begin{document}

\title[]{Closed-form existence conditions for band-gap resonances in a finite periodic chain under general boundary conditions}



\author{Mary V. Bastawrous}
\email{mary.bastawrous@colorado.edu}
\affiliation{Ann and H.J. Smead Aerospace Engineering Sciences, University of Colorado Boulder, Boulder, CO, 80303, USA}
\author{Mahmoud I. Hussein}
\email{mih@colorado.edu}
\affiliation{Ann and H.J. Smead Aerospace Engineering Sciences, University of Colorado Boulder, Boulder, CO, 80303, USA}










\begin{abstract}
Bragg scattering in periodic media generates band gaps, frequency bands where waves attenuate rather than propagate. Yet, a finite periodic structure may exhibit resonance frequencies within these band gaps. This is caused by boundary effects introduced by the truncation of the nominal infinite medium. Previous studies of discrete systems determined  existence conditions for band-gap resonances, although the focus has been limited to periodic chains with free-free boundaries. In this paper, we present closed-form existence conditions for band-gap resonances in discrete diatomic chains with general boundary conditions (free-free, free-fixed, fixed-free, or fixed-fixed), odd or even chain parity (contrasting or identical masses at the ends), and the possibility of attaching a unique component (mass and/or spring) at one or both ends. The derived conditions are consistent with those theoretically presented or experimentally observed in prior studies of structures that can be modeled as linear discrete diatomic chains with free-free boundary conditions. An intriguing case is a free-free chain with even parity and an arbitrary additional mass at one end of the chain. Introducing such an arbitrary mass underscores a transition among a set of distinct existence conditions depending on the type of the chain boundaries and parity. The proposed analysis is applicable to linear periodic chains in the form of lumped-parameter models as well as continuous granular media models examined in the low-frequency regime. 
\end{abstract}


\maketitle



\newcommand\bigzero{\makebox(0,0){\text{\huge0}}}

\section{Introduction}
Periodic media exhibit band gaps due to the Bragg scattering effect at spatially repeating interfaces. Underlying this phenomenon are destructive wave interferences at wavelengths comparable to the lattice constant \cite{brillouin1953wave,deymier2013acoustic,hussein2014}. 
However, a finite periodic structure (truncated periodic medium) may have resonance frequencies within the band gaps due to the introduction of boundaries \cite{wallis1957effect,camley1983transverse}.
The mode shapes associated with band-gap resonances are localized, exhibiting high amplitudes at the boundaries$-$causing what are widely known as \it surface modes\rm$-$and exponential decay inwards due to the band-gap attenuation effect. 
Band-gap resonances in finite periodic structures occur in different types of physical problems, e.g., elastic\cite{camley1983transverse,Boud1993surface,el1996theory,hladky2005localized,ren2007theory,hladky2007experimental,el2009acoustic,ren2010surface,theocharis2010intrinsic,davis2011analysis,hvatov2015free,al2017pole,Shuvalov2018,al2019dispersion}, electromagnetic \cite{yeh1978optical,djafari2001surface,vinogradov2006surface}, magnetic  \cite{puszkarski1979theory,ferchmin2001existence}, atomic-lattice-dynamics \cite{wallis1957effect,cheng1969vibration,puszkarski1983effect,wallis1994surface,allen2000surface}, and quantum-mechanics \cite{ren2006electronic} problems. 
The phenomenon is therefore relevant to a wide range of applications, including sensing \cite{li2013phase, bonhomme2019love}, transitional flow control \cite{hussein2015flow,Barnes_2021}, and quantum computing \cite{andrich2017long}, to name a few. Applications that utilize vibration localization in general, such as piezoelectric energy harvesting \cite{thorp2001attenuation,gonella2009interplay,Carrara_2013,lv2013vibration,Sugino2018}, would also benefit from the ability to engineer band-gap resonances.  

Despite their common appearance, band-gap resonances are not always guaranteed to arise upon truncation of a periodic medium. 
Their existence depends on certain factors, such as the truncation-point location and the type of boundary conditions \cite{ren2010surface}.  
In continuous elastic media with one-dimensional periodicity, a finite periodic structure comprised of an integer number of unit cells always has a band-gap resonance at one of its edges \cite{ren2007theory,el2009acoustic}. 

It has been proven that in a given band gap for a semi-infinite propagating medium, one band-gap resonance at most appears for symmetric configurations while a maximum of three appear for asymmetric configurations \cite{Shuvalov2018}.
However, alteration of the boundary conditions, e.g., to a fixed surface boundary, can alter these existence criteria \cite{ren2007theory,ren2010surface,hvatov2015free,Shuvalov2018}. 
The behaviour of band-gap resonance frequencies upon adding a variable-thickness cap layer is another feature that has been investigated, theoretically, numerically, and experimentally \cite{el2009acoustic,davis2011analysis}. 
The relationship between band-gap resonances under certain boundary conditions and the natural frequencies of the constituent unit cell under the same boundary conditions has been established for both symmetric and asymmetric configurations in engineering structures, such as plates and beams \cite{hvatov2015free}.

Band-gap resonances in discrete periodic structures exhibit trends that differ significantly from those in continuous periodic structures. 
For instance, a discrete one-dimensional periodic structure can feature as many band-gap resonances as the number of masses in its unit cell \cite{wallis1957effect,cheng1969vibration}, as opposed to its continuous counterpart$-$a periodic layered medium$-$that does not depend on the number of unit-cell layers  \cite{el1996theory}. 
Moreover, the existence conditions for band-gap resonances in a discrete periodic structure may depend on the finite number of its unit cells \cite{wallis1957effect,hladky2005localized, hladky2007experimental}, as opposed to the case in continuous structures where there is no dependency on the number of unit cells \cite{el2009acoustic,hvatov2015free, Shuvalov2018}.
Generally, a discrete model is more representative than a continuous one when particles can be seen as rigid bodies vibrating around their equilibrium positions. 
Some examples are atoms in a crystal \cite{wallis1994surface}, glued or welded beads vibrating mostly as rigid bodies at low frequencies \cite{hladky2005localized, hladky2007experimental}, and granular chains \cite{theocharis2010intrinsic}. 
Several studies have investigated band-gap resonances in such discrete structures \cite{wallis1957effect, cheng1969vibration, puszkarski1983effect, allen2000surface}. Wallis studied band-gap resonance frequencies in discrete free-free chains that are built from diatomic unit cells having different masses but similar interconnecting spring constants \cite{wallis1957effect}. 
The band-gap resonances' existence conditions were derived using Rutherford continuants \cite{rutherford1948xxv}. 
Later, H. Puszkarski studied the effect of different boundary force constants on the existence of band-gap resonances in a diatomic chain \cite{puszkarski1983effect}. 
This study presents the only attempt known to the authors to tackle the effect of boundary conditions on band-gap resonances' existence conditions in discrete structures; however, it is not generalizable to the possibility of attaching a unique component at the chain ends (e.g,, boundary masses that are distinct from the unit-cell masses), and does not incorporate diatomic unit cells having different spring constants.  
More recently, Al Ba'ba' et al. \cite{al2017pole,al2019dispersion} solved for the band-gap resonance frequencies of a free-free even chain by an approach that allowed for the possibility of periodic unit cells having different spring constants. The frequency expressions contained the existence conditions by default, and were obtained using a scheme developed by Da Fonseca to derive the characteristic equations of these systems \cite{Fonseca2007char}. 
In the case of equal spring constants in the diatomic unit cell, the frequency expressions reduced to those derived by Wallis  \cite{wallis1957effect}.
However, these could only be obtained if the characteristic equation could be explicitly solved, which is limited to a few exceptional cases such as a free-free even chain or a fixed-fixed odd chain.  \\
\indent In this paper, we formulate closed-form existence conditions for band-gap resonances in a periodic chain for  general boundary conditions, chain parity, and allowing for the possibility of attaching unique components at the chain ends. This wide range of different possibilities will make the analysis relevant to a broader range of applications. Knowledge of how the properties of band-gap resonances depend on the unit-cell parameters, the  global structure boundary conditions, the chain parity, and the properties of an altered mass and/or spring at the ends (when needed) provides a powerful tool to engineer and control these resonances. Owing to their localized modal nature at the edge or surface and their level of separation from other pass-band resonance frequencies, band-gap resonances offer unique design opportunities in applications that particularly require at least one of these traits$-$such as in certain sensing \cite{bonhomme2019love} applications and in flow control by phononic subsurfaces \cite{hussein2015flow,Barnes_2021}. 

%

\section{General  formulation: Equations of motion and existence conditions}
\label{sec:Approach}
In this section, we develop our general formulation for determining  existence conditions for band-gap resonances in discrete periodic chains comprised of a finite number of diatomic unit cells. We consider the possibility of a global arrangement that follows either an odd or even parity, and consider the four possible types of boundary conditions, namely, free-free, free-fixed, fixed-free, and fixed-fixed. Furthermore, we allow for the spring or mass at each boundary to be arbitrary, i.e., not necessarily the same as the springs and masses in the unit cell. 
The approach exploits the characteristic-equation derivation scheme developed by Da Fonseca \cite{Fonseca2007char} and followed by Al Ba'ba' et al. \cite{al2017pole,al2019dispersion}. 
While we focus on a diatomic unit cell, extension to an arbitrary number of masses in the unit cell is possible, as in the parent scheme of Da Fonseca \cite{Fonseca2007char,al2019dispersion}.
First, we present the problem of a diatomic chain with different boundary conditions and parity. Then, we provide a brief review of the Da Fonseca scheme and apply it to the derivation of the underlying characteristic equations. Finally, we derive the existence conditions for the band-gap resonances.  

\begin{figure*}[htbp!]
\centering
\centering
\includegraphics{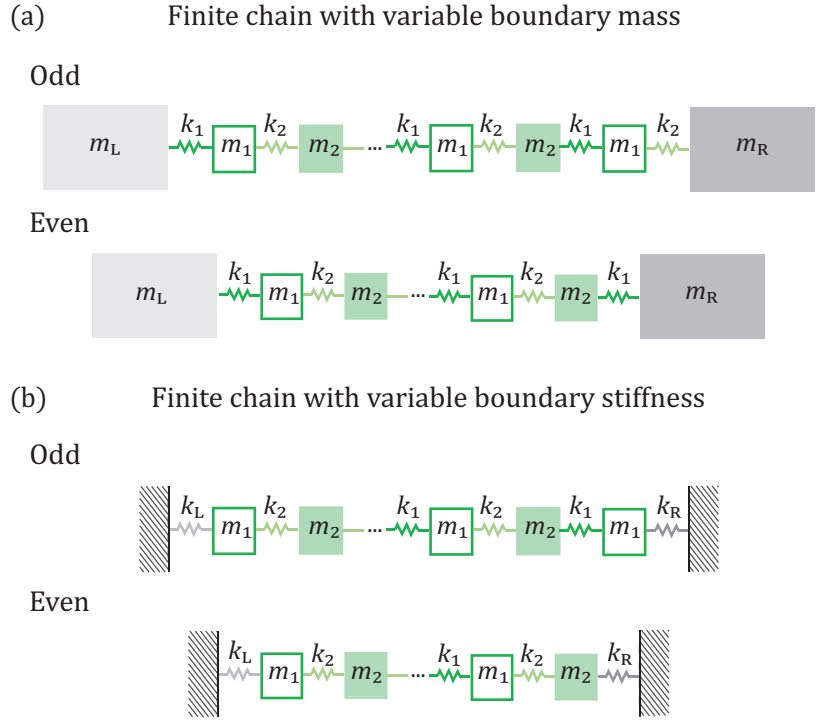}  
\caption{\label{fig:introX}{
Finite periodic chains with odd and even parity as well as different left and right boundary conditions. The boundary \textbf{(a)} masses and  \textbf{(b)} springs may also take on arbitrary values. In \textbf{(b)}, the boundary springs are attached to a rigid wall}.}          
\end{figure*}

\subsection{Characteristic Equations}

Consider the diatomic chains depicted in \autoref{fig:introX} which consist of $2n+1$ or $2n$ masses for the odd or even chains, respectively. Note the different diatomic-unit-cell masses and springs in \autoref{fig:introX}, colored in light and dark green.
The chain boundaries are modeled as either free boundary masses [case (a) in \autoref{fig:introX}(a)] or as springs connected to a rigid wall [case (b) in \autoref{fig:introX}(b)]. 
The boundary masses and spring constants can be altered as neeeded for any given application, and do not need to match those in the diatomic unit cell. In case (a) [\autoref{fig:introX}(a)], the function $u_\text{L} (t)$ denotes the displacement of mass $l$ in the chain, where $l = 0, 1, ... , 2n+2$ for the odd chain and $l = 0, 1, ..., 2n+1$ for the even chain. The masses are numbered such that the odd chain consists of $n + {1}/{2}$ diatomic unit cells and the even chain consists of $n$ diatomic unit cells. 
The equations of motion are written as 
\begin{subequations}
\begin{align}
m_\text{L} \ddot{u}_0 + k_2 (u_0 - u_1) & = 0,  \\
m_2 \ddot{u}_\text{L} + k_2 (u_\text{L} - u_{l-1}) + k_1 (u_\text{L} - u_{l+1}) & =0, \quad \textrm{even} \, \, l\\
m_1 \ddot{u}_\text{L} + k_2 (u_1 - u_{l+1}) + k_1(u_\text{L} - u_{l-1})  &= 0, \quad \textrm{odd} \, \, l\\
m_\text{R} \ddot{u}_{2n+2} + k_2(u_{2n+2} - u_{2n+1})  &=0, \quad \textrm{odd-chain right boundary}, \\
m_\text{R} \ddot{u}_{2n+1} + k_1(u_{2n+1} - u_{2n}) &=0, \quad \textrm{even-chain right boundary}.
\end{align}
\label{eq:chainA}
\end{subequations}
It is seen that the right-boundary equation changes depending on the the chain parity.

On the other hand in case (b) [\autoref{fig:introX}(b)], we have the index $l = 1, ... , 2n+1$ for the odd chain and the index $l = 1, ..., 2n$ for the even chain. Here, the equations of motion are written as
\begin{subequations} 
\begin{align}
m_1 \ddot{u}_1 + k_2 (u_1 - u_2) + \kappa_\text{L} u_1 & = 0,  \\
m_2 \ddot{u}_\text{L} + k_2 (u_\text{L} - u_{l-1}) + k_1 (u_\text{L} - u_{l+1}) & =0, \quad \textrm{even}\, \, l\\
m_1 \ddot{u}_\text{L} + k_2 (u_1 - u_{l+1}) + k_1(u_\text{L} - u_{l-1})  &= 0, \quad \textrm{odd} \, \, l\\
m_1 \ddot{u}_{2n+1} + k_1(u_{2n+1} - u_{2n}) + \kappa_\text{R} u_{2n+1} &=0, \quad \textrm{odd-chain right boundary},\\
m_2 \ddot{u}_{2n} + k_2(u_{2n} - u_{2n-1}) + \kappa_\text{R} u_{2n} &=0, \quad \textrm{even-chain right boundary}.
\label{eq:chainB}
\end{align}
\end{subequations}
Note that the equations of motion for the internal masses match those derived for case (a) and written in Eqs. \eqref{eq:chainA} for odd and even mass indexes. 

For both boundary-condition cases, the governing equations of motion can be expressed in matrix form as 
\begin{equation}
\bold{M} \, \ddot{\bold{u}} + \bold{K} \, \bold{u} = \bold{0},
\label{eq:periodicEOM}
\end{equation}
where $\bold{M}$ and $\bold{K}$ are the mass and stiffness matrices, respectively, and $\bold{u}$ is the nodal degrees-of-freedom vector. 
Assuming harmonic motion, we set $u_\text{L} = q_\text{L} \exp{(i \omega t)} $. Then, Eq. \eqref{eq:periodicEOM} becomes
\begin{gather} 
\bold{K_\text{d}} \, \bold{q} = \bold{0} , \quad \bold{K_\text{d}} = \bold{K} - \omega^2 \bold{M},
\label{eq:EVP}
\end{gather}
where $\bold{K_\text{d}}$ is the dynamic-stiffness matrix and $\bold{q}$ is the state vector.
Eq. \eqref{eq:EVP} is an eigenvalue problem for which the eigenvalues $\omega^2$ can be obtained by setting the determinant of $\bold{K_\text{d}}$ to zero, 
\begin{equation}
|\bold{K_\text{d}}| = 0,
\end{equation}
and solving for the roots of the resulting characteristic equation. 
Using a normalized frequency $\Omega$ such that 
\begin{equation}
\Omega = \omega/\sqrt{k_1/m_1},
\end{equation}
unit-cell parameters
\begin{gather}
m_f = \frac{m_2}{m_1}, \quad \textrm{and } k_f = \frac{k_2}{k_1},
\end{gather}
and boundary parameters
\begin{gather}
\mu_\text{L} = \frac{m_\text{L}}{m_1}, \quad \mu_\text{R} = \frac{m_\text{R}}{m_1}, \quad \kappa_\text{L} = \frac{k_\text{L}}{k_1}, \quad \textrm{and } \kappa_\text{R} = \frac{k_\text{R}}{k_1}, 
\end{gather}
the odd-chain normalized dynamic-stiffness matrix can be written in the form
\begin{eqnarray}
\bold{K_\text{d}}  = \left(
\begin{array}{cccccc}
d_1 + \alpha & a_1 \\
a_1 & d_2 & a_2 & & \text{\huge0} & \\ 
& a_2 & d_1 & a_1  & \\
 \text{\huge0} &  & \ddots \quad &  \ddots \quad  \ddots &\\
& & a_2 & d_2 & a_1 \\
&  & & a_1 & d_1 + \beta 
    \end{array}
 \right),
 \label{eq:OddKd}
\end{eqnarray}
where the values for $d_1$, $d_2$, $a_1$, $a_2$, $\alpha$ and $\beta$ are tabulated in \autoref{Tab:ParameterValues} for the different boundary-condition cases for the odd chains in \autoref{fig:introX}.
For the even-chain, the normalized dynamic-stiffness matrix takes the same form as in Eq. \eqref{eq:OddKd}, except for the omission of the last row and column and adding $\beta$ to the last diagonal term of the remaining matrix to become $d_2 + \beta$. 
The values corresponding to $d_1$, $d_2$, $a_1$, $a_2$, $\alpha$, and $\beta$ for the different boundary-condition cases for the even chains in \autoref{fig:introX} are also tabulated in \autoref{Tab:ParameterValues}.\\
\indent Upon inspection, it is clear that the dynamic-stiffness matrices, such as the one shown in Eq. \eqref{eq:OddKd}, are similar in form to matrices known as perturbed 2-Toeplitz matrices \cite{Fonseca2007char,al2017pole}.  
This similarity is due to the fact that matrices have two periodically repeating values $d_1$ and $d_2$ along the diagonals, and two repeating values $a_1$ and $a_2$ along the sub- and super- diagonals. 
Note that these four parameters pertain to the diatomic unit cell.
Moreover, the first and last diagonal elements are perturbed by the parameters $\alpha$ and $\beta$, which pertain to the properties of the end springs and masses and applied boundary conditions, hence the term `perturbed' 2-Toeplitz matrices. 
Da Fonseca showed that the characteristic equations for perturbed 2-Toeplitz matrices can be obtained as \cite{Fonseca2007char}
\begin{equation}
\Delta_{2m+1} = \left(-d_1 - \alpha - \beta\right) P_m(\pi(\Omega)) - \left(\alpha \beta d_2 + \alpha a_1^2 - \beta a_2^2\right)  P_{m-1}(\pi(\Omega)) 
\label{eq:oddChar}
\end{equation}
for a problem with an odd matrix dimension, and 
 \begin{equation}
\Delta_{2m} =  P_m(\pi(\Omega)) - \left( \alpha d_2 + \beta d_1 + \alpha \beta + a_2^2\right)  P_{m-1}(\pi(\Omega))  + \alpha \beta a_1^2 P_{m-2}(\pi(\Omega))
\label{eq:evenChar}
\end{equation}
for a problem with an even matrix dimension, where 
\begin{subequations}
\begin{eqnarray}
\pi(\Omega) = d_1 d_2, \\
P_m(x) = (a_1 a_2)^m  U_m\left(x \right),
\end{eqnarray}
\end{subequations}
and where $U_m(x)$ is a Chebyshev polynomial of the second kind,
\begin{subequations}
\begin{gather}
U_m(x) = \frac{\sin\ (m+1) \theta }{\sin \theta},  \quad\\
x = \cos\theta =\left(\frac{\pi(\Omega) - a_1^2 - a_2^2}{2 a_1 a_2}\right).\label{eq:argument}
\end{gather}
\end{subequations}
Note that if the dimension of the dynamic-stiffness matrix is $2m+1$ for the odd-matrix problem and $2m$ for the even-matrix problem, $m$ does not necessarily correspond to the number of diatomic unit cells $n$. For example, $m = n+1$ in case (a) in \autoref{fig:introX}(a), and $m = n$ in case (b) in \autoref{fig:introX}(b). 
\begin{table}[hb!]
\centering
\begin{minipage}{40em}
    \caption{Parameter values for odd and even chains for the cases in \autoref{fig:introX}(a) and (b)}
    \centering
      \setlength{\tabcolsep}{3pt}
  \begin{tabular}{c|cccccc}
   \textbf{Cases} & $d_1$ & $d_2$ & $a_1$ & $a_2$ & $\alpha$ & $\beta$ \\ \hline\hline
    Odd (a) & \multirow{2}{*}{$1+k_f - m_f \Omega^2$} & \multirow{2}{*}{$1+ k_f-\Omega^2$ } & \multirow{2}{*}{$-k_f$}  & \multirow{2}{*}{$-1$} & \multirow{2}{*}{$\Omega^2 (m_f - \mu_\text{L}) - 1$}  & $\Omega^2(m_f - \mu_\text{R})-1$\\ 
    Even (a) & & & & & & $\Omega^2(1 - \mu_\text{R})-k_f$ \\\hline
   Odd (b) & \multirow{2}{*}{$1+ k_f-\Omega^2$ } & \multirow{2}{*}{$1+k_f - m_f \Omega^2$ }& \multirow{2}{*}{$-1$}  &  \multirow{2}{*}{$-k_f$} &  \multirow{2}{*}{$\kappa_\text{L} - 1$} & $\kappa_\text{R} - k_f$ \\ 
   Even (b) &  & & & & & $\kappa_\text{R} -1$
    \label{Tab:ParameterValues}
    \end{tabular}   
\end{minipage}
\end{table}


Using this approach, it can be shown that the characteristic equation generally takes the form 
\begin{equation}
A \left(c_2 \sin(n+2)\theta + c_1 \sin(n+1)\theta + c_0 \sin n\theta\right) = 0, 
\label{eq:charGen}
\end{equation}
where $A = A(\theta,\Omega, k_f, m_f)$, $c_i = c_i (\Omega, k_f,m_f,g_\text{L},g_\text{R}), i=$ 0, 1 and 2. 
These coefficients are tabulated in \autoref{Tab:CharCoeffValues} for different boundary-condition cases and chain parity. 
Notice that a free $\Omega^2$ term, which represents rigid-body motion in case (a), is present only in the odd-chain characteristic equation. 
Nevertheless, it shows in the even-chain characteristic equation when taking the limit as $\theta \to 0$, which represents the long-wavelength limit.

This concludes the discussion on the utilization of the Da Fonseca scheme to formulate the characteristic equations for diatomic chains with different parity, boundary components, and boundary conditions. 
In the next subsection, we extract the existence conditions for band-gap resonances using the derived characteristic equations. 

\subsection{Existence Conditions for Band-gap Resonance Frequencies}
  
\begin{figure}[htbp!]
\centering
\centering
\includegraphics{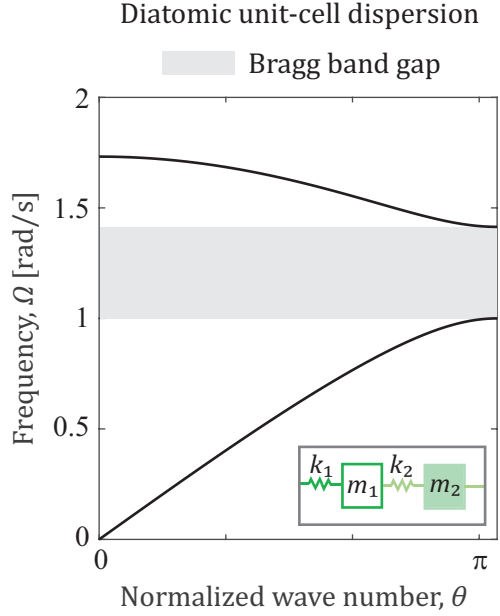}
\caption{Dispersion curves for a diatomic unit cell (shown in the inset) where $m_1 = 1, k_1 = 1, m_2 = 2, k_2 = 1$. The Bragg band gap between the acoustic and optical branches is shaded in grey.}      
\label{fig:dispersion}     
\end{figure}

Looking at the input argument for the Chebyshev polynomial of the second kind in Eq. \eqref{eq:argument}, it can be rearranged as 
\begin{equation}
m_f \Omega^4 - (1+k_f)(1+m_f) \Omega^2 + 2 k_f (1- \cos \theta) = 0.
\label{eq:diatomicDispersion}
\end{equation}
This relation (plotted in \autoref{fig:dispersion}) is in fact the diatomic unit-cell dispersion relation \cite{deymier2013acoustic,hussein2014}.
Note that $\theta$ here corresponds to the normalized wave number. 
Given the nature of a dispersion relation, both Eqs. \eqref{eq:argument} and \eqref{eq:diatomicDispersion} contain parameters that are only relevant to the unit cell, regardless of the applied boundary conditions on a truncated finite version of the chain. 
Solving for the roots of Eq. \eqref{eq:diatomicDispersion} we obtain
\begin{equation}
\Omega^2_{1,2}  = \frac{(1+k_f)(1+m_f) \pm \sqrt{(1+k_f)^2(1+m_f)^2 - 8 k_f m_f (1-\cos \theta)}}{2 m_f},
\label{eq:OmegaSquared}
\end{equation}
which yields direct expressions for the dispersion acoustic and optical branches. The acoustic and optical frequencies at the edge of the Irreducible-Brillouin-Zone (IBZ) edge are obtained by setting $\theta = \pi$ in Eq. \eqref{eq:OmegaSquared}, which gives
\begin{equation}
\Omega^2_{1,2}  = \frac{(1+k_f)(1+m_f) \pm \sqrt{(1+k_f)^2(1+m_f)^2 - 16 k_f m_f}}{2 m_f}.
\label{eq:IBZedgeSquared}
\end{equation}
The edge frequencies are the positive square-roots of Eqs. \eqref{eq:IBZedgeSquared} as the negative roots are discarded. 
Having established this connection between the unit-cell dispersion and the argument of the Chebeyshev polynomial of the second kind,  we turn back to the characteristic equation [Eq. \eqref{eq:charGen}], which can be generally written as 
\begin{equation}
\Delta \left(\theta\left(m_{f}, k_{f}, \Omega\right), \Omega, k_{f}, m_{f}, n, g_{L}, g_{R}\right)=0,
\end{equation}
where $g_\text{L}$ and $g_\text{R}$ are general parameters that represent the left and right boundary conditions. 
Taking the limit of the characteristic equation as $\theta \to \pi$ and using L'h\^{o}pital's rule,
\begin{equation}
\lim _{\theta \rightarrow \pi} \Delta \left(\theta\left(m_{f}, k_{f}, \Omega\right), \Omega, k_{f}, m_{f}, n, g_{L}, g_{R}\right)=0,
\label{eq:LimCharEqn}
\end{equation}
we combine Eq. \eqref{eq:LimCharEqn} with Eqs. \eqref{eq:IBZedgeSquared} (both the acoustic and optical branches' squared frequencies) to obtain closed-form expressions in the form
\begin{equation}
f\left(k_{f}, m_{f}, n, g_{L}, g_{R}\right)=0.
\label{eq:BorderEq}
\end{equation}

\begin{figure}[htbp!]
\centering
\centering
\includegraphics{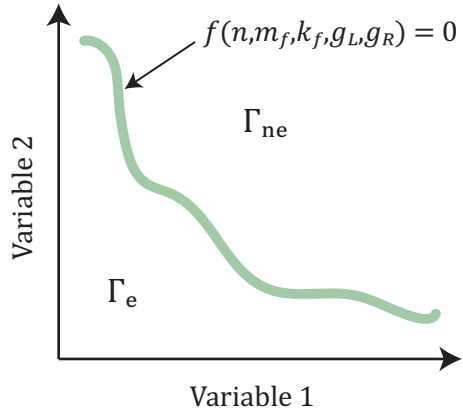}  
\caption{Schematic of a two-dimensional functional relationship indicating regions of band-gap resonance existence and nonexistence in terms of any two independent variables from among those listed in the argument of the \it f\rm-function.}   
\label{fig:Schematic}        
\end{figure}

Each one of these closed-form expressions is a hyper-surface equation that marks the border surface corresponding to the first IBZ edge. This can be justified by the fact that taking the characteristic-equation limit as $\theta \to \pi$ represents approaching the edges of the IBZ, which bounds the band gaps. Thus, these border surfaces differentiate the existence regions$-$defined in terms of the value intervals of the finite-chain parameters$-$of the band-gap resonance frequencies from other regions where they do not exist.  
As the finite-chain parameters are perturbed from their values at the IBZ edge in the direction of their existence regions, band-gap resonances result from either an acoustic-branch frequency crossing into the Bragg band gap (from the lower edge), or an optical-branch frequency crossing into the Bragg band gap (from the upper band-gap edge). 
The 2D schematic in \autoref{fig:Schematic} illustrates this idea for a single band-gap resonance frequency in which the hyper-surface in Eq. \eqref{eq:BorderEq} is reduced to a 2D curve by fixing all the parameters except for two. 
Each point on this curve represents a crossing point from regions of non-existent band-gap resonances $\Gamma_{\textrm{ne}}$ to regions of existent band-gap resonances $\Gamma_{\textrm{e}}$. 
Note that the normalized wave number $\theta$ will be complex in existence regions (due to band-gap attenuation), whereas it will be real in nonexistence regions. Thus, these regions can be defined as the set of parameters that both satisfy the characteristic equation and result in the suitable wave-number $\theta$,  
\begin{subequations}
\begin{gather}
\Gamma_{\text {e}}:=\left\{\left(m_{f}, k_{f}, n, g_{L}, g_{R}\right): \Delta \left(\theta\left(m_{f}, k_{f}, \Omega\right), \Omega, k_{f}, m_{f}, n, g_{L}, g_{R}\right)=0, \theta\left(k_{f}, m_{f}\right) \in \mathbb{C}\right\}, \\
\Gamma_{\text {ne}}:=\left\{\left(m_{f}, k_{f}, n, g_{L}, g_{R}\right): \Delta \left(\theta\left(m_{f}, k_{f}, \Omega\right), \Omega, k_{f}, m_{f}, n, g_{L}, g_{R}\right)=0, \theta\left(k_{f}, m_{f}\right) \in \mathbb{R}\right\}.
\label{eq:Regions}
\end{gather}
\end{subequations}
Therefore, testing whether $\theta \in \mathbb{C}$ or not in the regions lying on either side of the hyper-surface in Eq. \eqref{eq:BorderEq} allows us to identify whether a particular region associated with existent or non-existent of band-gap resonances. 

Applying this procedure to the characteristic-equation general template in Eq. \eqref{eq:charGen} and taking its limit  as $\theta \to \pi$, yields an equation of the form
\begin{gather}
b_0 \Omega^4 + b_1 \Omega^2 + b_2 = 0, 
\label{eq:BorderInter}
\end{gather}
in which $\Omega^2$ can be replaced by either the acoustic- or the optical-branch IBZ-edge frequency in Eq. \eqref{eq:IBZedgeSquared}. 
This yields the hyper-surface expression in Eq. \eqref{eq:BorderEq} for each branch crossing. 
The constants $b_0$, $b_1$ and $b_2$ are tabulated in \autoref{Tab:ExisCoeffValues} for the different boundary-condition and parity cases in \autoref{fig:introX}.

\begin{sidewaystable}[hb!]
\centering
\begin{minipage}{57em}
    \caption{Coefficients of the characteristic equations [Eq. \eqref{eq:charGen}] for odd and even chains: Cases in \autoref{fig:introX}(a) and (b)}
    \centering
      \setlength{\tabcolsep}{3pt}
  \begin{tabular}{c|cccc}
   \textbf{Cases} & $A$ & $c_2$ & $c_1$ & $c_0$ \\ \hline\hline
   \multirow{3}{*} {Odd-(a)} & \multirow{3}{*} {$\csc(\theta) \Omega^2$} &\multirow{3}{*} { $k_f (\mu_\text{L} + \mu_\text{R} - m_f)$} & $\Omega^4 (m_f - \mu_\text{L})(m_f - \mu_\text{R}) + \Omega^2 \left[ m_f (1+k_f) \right.$  & \multirow{3}{*} {$0$}\\
   & & & $\left. (1+m_f-\mu_\text{L}) + \mu_\text{L} + \mu_\text{R} (k_f + m_f + k_f m_f \right.$ & \\
   & & & $\left.-(1+k_f) \mu_\text{L}) \right] +  k_f \left[1 + 2 m_f - (\mu_\text{L} + \mu_\text{R})\right]$ &\\
    
   \multirow{2}{*} { Even-(a)}& \multirow{2}{*} {$\csc(\theta)$} &\multirow{2}{*} { $k_f^2$} & $k_f \left[ \Omega^4 \left(\mu_\text{L} \mu_\text{R}-m_f \right) + \Omega^2 \right.$ &   $\Omega^4 (m_f - \mu_\text{L})(1 - \mu_\text{R})+ k_f \Omega^2 $ \\
    & & & $\left.\left(1 + k_f + m_f( 1+ k_f) - \mu_\text{L} - \mu_\text{R} \right) -2 k_f^2 \right]$ & $\left(\mu_\text{L} + \mu_\text{R}-(1 + m_f) \right) + k_f^2$ \\
     \hline
     
\multirow{2}{*} {Odd-(b)}& \multirow{2}{*} {$\csc(\theta)$} &$k_f \left( \Omega^2 - (\kappa_\text{L} + \kappa_\text{R})\right)$  &  \multirow{2}{*} {$m_f \Omega^2 (1-\kappa_\text{L})(k_f - \kappa_\text{R})$} & \multirow{2}{*} {$0$}\\ 
& &$+ k_f (\kappa_\text{L} + \kappa_\text{R}) - \kappa_\text{L} \kappa_\text{R} (1+k_f)$  &\\

\multirow{2}{*} {Even-(b)} & \multirow{2}{*} {$\csc(\theta)$}   &  \multirow{2}{*} {$k_f^2$} & $k_f \left[\Omega^2 (1 - \kappa_\text{R} + m_f(1 - m_f))+\right.$ & \multirow{2}{*} {$k_f^2(1 - \kappa_\text{L})(1 - \kappa_\text{R})$}  \\
   & & & $\left.(\kappa_\text{L} \kappa_\text{R} + k_f (\kappa_\text{L} + \kappa_\text{R}-2))\right]$ & 
    \label{Tab:CharCoeffValues}
    \end{tabular}   
\end{minipage}
\hrule
\vspace{0.2in}

\begin{minipage}{57em}
    \caption{Coefficients of existence conditions for band-gap resonances [Eq. \eqref{eq:BorderInter}] for odd and even chains: Cases in \autoref{fig:introX}(a) and (b)}
    \centering
      \setlength{\tabcolsep}{3pt}
  \begin{tabular}{c|ccc}
   \textbf{Cases} & $b_0$ & $b_1$ & $b_2$ \\ \hline\hline
   \multirow{2}{*} {Odd-(a)} & \multirow{2}{*} {$-(1 + n) (m_f - \mu_\text{L}) (m_f - \mu_\text{R})$} & $(1 + n) \left[(1 + k_f)( m_f^2- m_f (\mu_\text{L} + \mu_\text{R} -1))\right.$ &  $k_f \left[3 (\mu_\text{L} + \mu_\text{R})+ 2 n (\mu_\text{L} + \mu_\text{R}) \right.$\\
   & &  $\left. - k_f \mu_\text{R}  + \mu_\text{L} (\mu_\text{R} +  k_f \mu_\text{R} -1)\right]$ & $\left. -(n+1) - m_f (4 + 3 n)\right]$\\
   
   \multirow{2}{*} {Even-(a)} & $ -n (m_f - \mu_\text{L}) (1-\mu_\text{R})$ & $k_f \left[k_f (1 + m_f) (1 + n)+ \right.$ &  \multirow{2}{*} { $- 4 k_f^2  (1 + n)$}\\
   & $- k_f (1 + n) (m_f - \mu_\text{L} \mu_\text{R})$ & $\left. (1 + 2 n) (1 + m_f - (\mu_\text{L} + \mu_\text{R}))\right]$
   \\\hline
   
    \multirow{2}{*} {Odd-(b)} & \multirow{2}{*} {0} & \multirow{2}{*} {$ -m_f(k_f - \kappa_\text{R})(1 - \kappa_\text{L})(2 + n) + k_f (3 + n)$} &  $\kappa_\text{L} \kappa_\text{R} (n+2) - k_f \kappa_\text{R} (2n+5)$ \\
    & & & $+ k_f \kappa_\text{L} (\kappa_\text{R} (n+2)-(2n+5))$\\
    
     \multirow{2}{*} {Even-(b)} & \multirow{2}{*} {0} & \multirow{2}{*} {$ k_f (1 -\kappa_\text{R} + m_f(1 - \kappa_\text{L}))(n+1)$} &  $ k_f \left[\kappa_\text{L} \kappa_\text{R} (n+1)+ k_f(\kappa_\text{L}+\kappa_\text{R}\right. $ \\
    & & & $\left.-4-n( \kappa_\text{L}-2)(\kappa_\text{R}-2))\right]$
    \label{Tab:ExisCoeffValues}
    \end{tabular}   
\end{minipage}
\end{sidewaystable}

Thus, we obtain the conditions of existence for band-gap resonance frequencies for different chain boundary conditions, unit-cell number, and other diatomic-unit-cell parameters.
Note that the different boundary conditions utilized in the two cases in \autoref{fig:introX} encompass different situations.
For instance, we can set the boundary mass in \autoref{fig:introX}(a) to zero to recover a free-boundary case or take the limit as it approaches infinity to recover a fixed-boundary case. 
Moreover, parameter values corresponding to certain conditions$-$or their asymptotic limits as other parameters go to infinity$-$can be obtained, which is useful in the design process for applications that rely on the existence or absence of band-gap resonances.
Furthermore, in the case of the diatomic chain, the roots of Eq. \eqref{eq:diatomicDispersion} yield only two band-gap resonance frequencies. 
However, if the unit cell has $k$ masses, the right-hand-side polynomial in Eq. \eqref{eq:argument} yields $k$ roots $\Omega_i^2$, where $i = 1, ..., k$. 
Therefore, this leads to the possibility of up to $k$ band-gap resonance frequencies. 
This conclusion agrees with the findings of Y. C. Cheng \cite{cheng1969vibration} who studied unit cells with quasi-periodicity. 

To conclude, this section shows a systematic approach that yields the closed-form existence conditions of band-gap resonances for diatomic chains with different chain parity and boundary conditions including the possibility of altering the boundary springs/masses.
In the upcoming sections, we will use some selected examples to demonstrate this approach and reveal the effect of boundary conditions, parity, and composition of boundary springs/masses on the existence of band-gap resonances. 

\section{Example of basic structure: Finite chain with free-free, free-fixed, fixed-free, or fixed-fixed boundary conditions}
\label{sec:TraditionalBoundaries}

The characteristic equations for both odd and even chains with a fixed or free boundary condition at each end, also reported by Al Ba'ba' et al. \cite{al2017pole}, are tabulated in \autoref{Tab:CharEqns}. 
These equations are obtained from the general characteristic equation by setting the boundary masses in the coefficients (\autoref{Tab:CharCoeffValues}) to zero in the free-boundary case, and by taking the coefficients' limit as the boundary masses approach infinity in the fixed-boundary case. 
Another approach is to directly consider the corresponding governing equations with the desired boundary conditions already Incorporated and obtain their characteristic equations.
Only two cases yield explicit frequency solutions: a free-free even chain and a fixed-fixed odd chain.
In the case of a free-free even chain, the resonance frequencies are solved for to give us 
\begin{subequations}
\begin{gather}
\Omega_1^2 = 0, \quad \Omega_{l+1,2n-1+1}^2: \theta_\text{L}(\Omega) = \frac{l \pi}{n}, \, l = 1, .. ,n-1, \label{eq:EvenFF_Pass}\\ 
\Omega^2_{n+1} = \frac{k_f (1 + m_f)}{m_f}. 
\label{eq:EvenFF_Stop}
\end{gather}
\end{subequations}
Note that Eq. \eqref{eq:EvenFF_Pass} can be used with Eq. \eqref{eq:argument} to solve for the pass-band resonance frequencies (as shown by Al Ba'ba' et al. \cite{al2017pole}), while Eq. \eqref{eq:EvenFF_Stop} represents that band-gap resonance frequencies. 
Similarly, the resonance frequencies in the case of a fixed-fixed odd chain are solved for to give us 
\begin{subequations}
\begin{gather}
\Omega_{l,2n+2-l}^2: \theta_\text{L}(\Omega) = \frac{l \pi}{n+1}, \, l = 1, .. ,n, \label{eq:OddXX_Pass} \\ 
\Omega^2_{n+1} = 1 + k_f, \label{eq:OddXX_Stop}
\end{gather}
\end{subequations}
where Eq. \eqref{eq:OddXX_Pass} represents the pass-band frequencies, while Eq. \eqref{eq:OddXX_Stop} represents the band-gap resonance frequency. \\
\indent Apart from these two cases, no other characteristic equations in \autoref{Tab:CharEqns} yield explicit solutions for the resonance frequencies, including band-gap resonances; thus, the characteristic equations need to be solved numerically for the frequency roots.
\begin{table}[hb!]
\centering
\begin{minipage}{40em}
    \caption{Characteristic equations for both odd and even diatomic chains with different boundary conditions, also reported in by Al Ba'ba' et al. \cite{al2017pole}}
    \centering
      \setlength{\tabcolsep}{4pt}
  \begin{tabular}{c|lc}
   \textbf{Parity} & \textbf{BC} & \textbf{Characteristic equation}  \\ \hline \hline
   \multirow{4}{*}{Odd} & Free-Free  & $\Omega^2 \csc\theta \left(m_f \sin n\theta + \sin(n+1)\theta \right) = 0$\\
   & Free-Fixed & $\csc \theta \left( k_f \sin n\theta + (\Omega^2-k_f) \sin (n+1)\theta \right) = 0$ \\
   & Fixed-Free & $\csc \theta \left( \sin n\theta + (\Omega^2-1) \sin (n+1)\theta \right) = 0$ \\ 
   & Fixed-Fixed & $\csc\theta \left(\Omega^2 - (1 + k_f) \right) \sin (n+1)\theta  = 0$\\
   \hline
   \multirow{4}{*}{Even} & Free-Free  &  $\Omega^2 \csc\theta \left(k_f (1+m_f) - m_f \Omega^2 \right) \sin n\theta  = 0 $\\ 
   & Free-Fixed & $\csc\theta \left((m_f \Omega^2 - k_f) \sin n\theta  + k_f \sin(n+1)\theta\right)  = 0$\\
    & Fixed-Free &$\csc\theta \left((\Omega^2 - k_f) \sin n\theta  + k_f \sin(n+1)\theta \right)  = 0$\\ 
   & Fixed-Fixed & $\csc\theta \left(\sin n\theta + k_f \sin(n+1)\theta \right)  = 0$ 
    \label{Tab:CharEqns}
    \end{tabular}   
\end{minipage}
\end{table}
However, the hyper-surfaces separating band-gap resonances' existence and non-existence regions for these diatomic chains can be obtained by following the approach in \autoref{sec:Approach}. 
As an example, we consider the free-fixed odd-chain case. 
Taking the limit of its characteristic equation in \autoref{Tab:CharEqns} yields
\begin{equation}
k_f (2n+1) - \Omega^2 (n+1) = 0.
\end{equation}
Using Eq. \eqref{eq:IBZedgeSquared}, the hyper-surface equations can be written as 
\begin{subequations}
\begin{gather}
k_f (2n+1) - \frac{n+1}{2m_f}\left((1+k_f)(1+m_f)-\sqrt{(1+k_f)^2(1+m_f)^2-16 k_f m_f}\right) = 0, \label{eq:OddFX_AcousHyp}\\ 
k_f (2n+1) - \frac{n+1}{2m_f}\left((1+k_f)(1+m_f)+\sqrt{(1+k_f)^2(1+m_f)^2-16 k_f m_f}\right) = 0,
\end{gather}
\label{eq:FX_Hyp}
\end{subequations}
for the acoustic- and optical-branch-crossing hyper-surfaces, respectively.  Testing the values of $\theta$ on either side of the hyper-surface equation allows us to conclude on which side the band-gap resonance frequencies exist. 
Moreover, solving for $k_f$ in either of Eqs. \eqref{eq:FX_Hyp} yields a rational function that includes both the acoustic- and optical-branch hyper-surfaces, as shown in \autoref{Tab:ExistenceConditions}. 
This rational function has a singularity point at $m_f = {(n+1)}/{n}$, which separates the hyper-surfaces of the acoustic and optical branches. 

\begin{figure*}[htbp!]
\centering
\centering
        \includegraphics{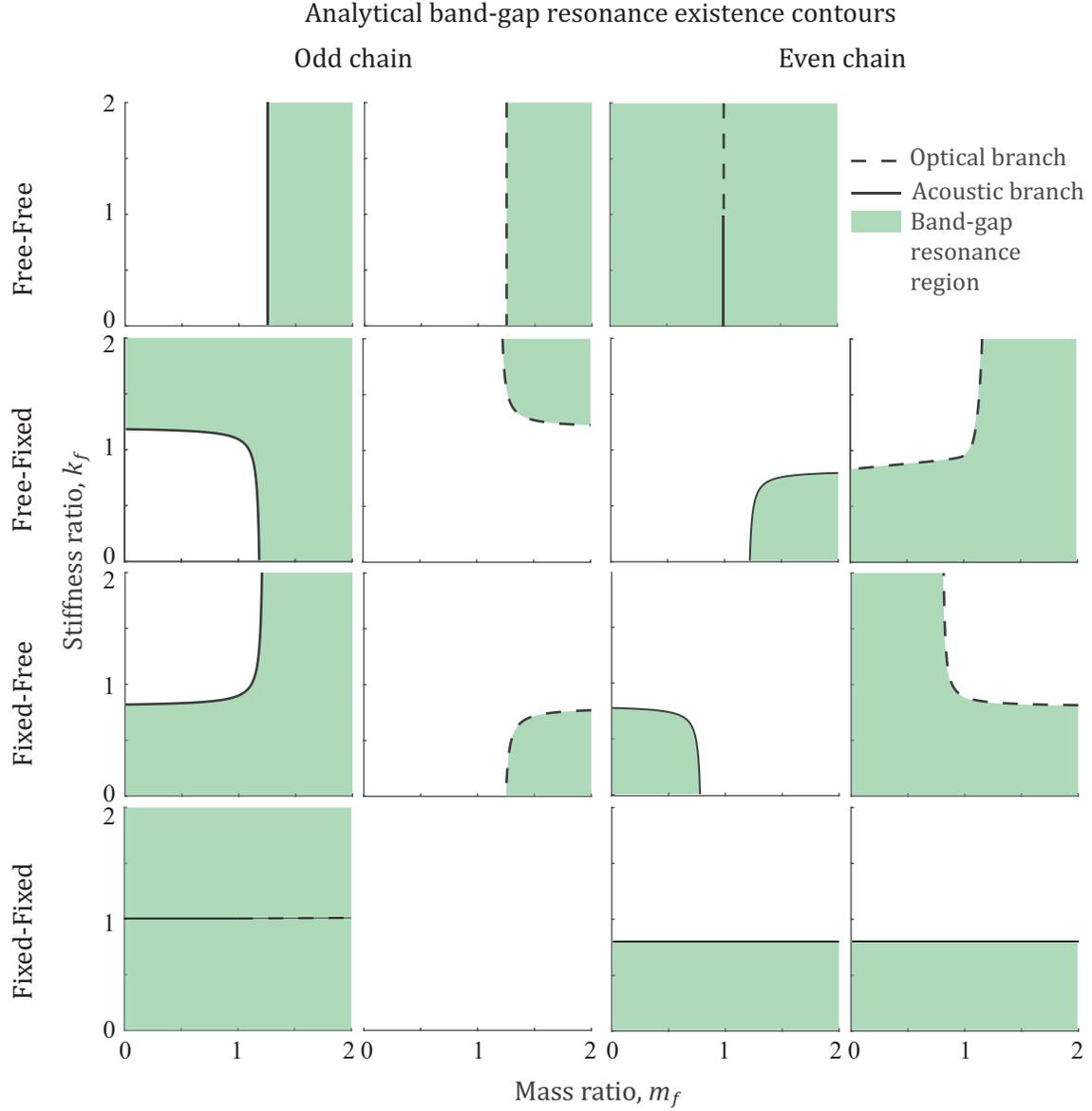}
             \caption{Analytical band-gap resonances existence contours for 5-unit-cell odd and even chains with free-free, free-fixed, fixed-free and fixed-fixed boundaries as a function of the diatomic unit cell mass and stiffness ratios $m_f$ and $k_f$. Acoustic branch crossing points are in solid lines, and optical branch crossing points are in dashed lines. }
        \label{fig:AnalRegions}
\end{figure*}

\begin{figure*}[htbp!]
\centering
\centering
        \includegraphics{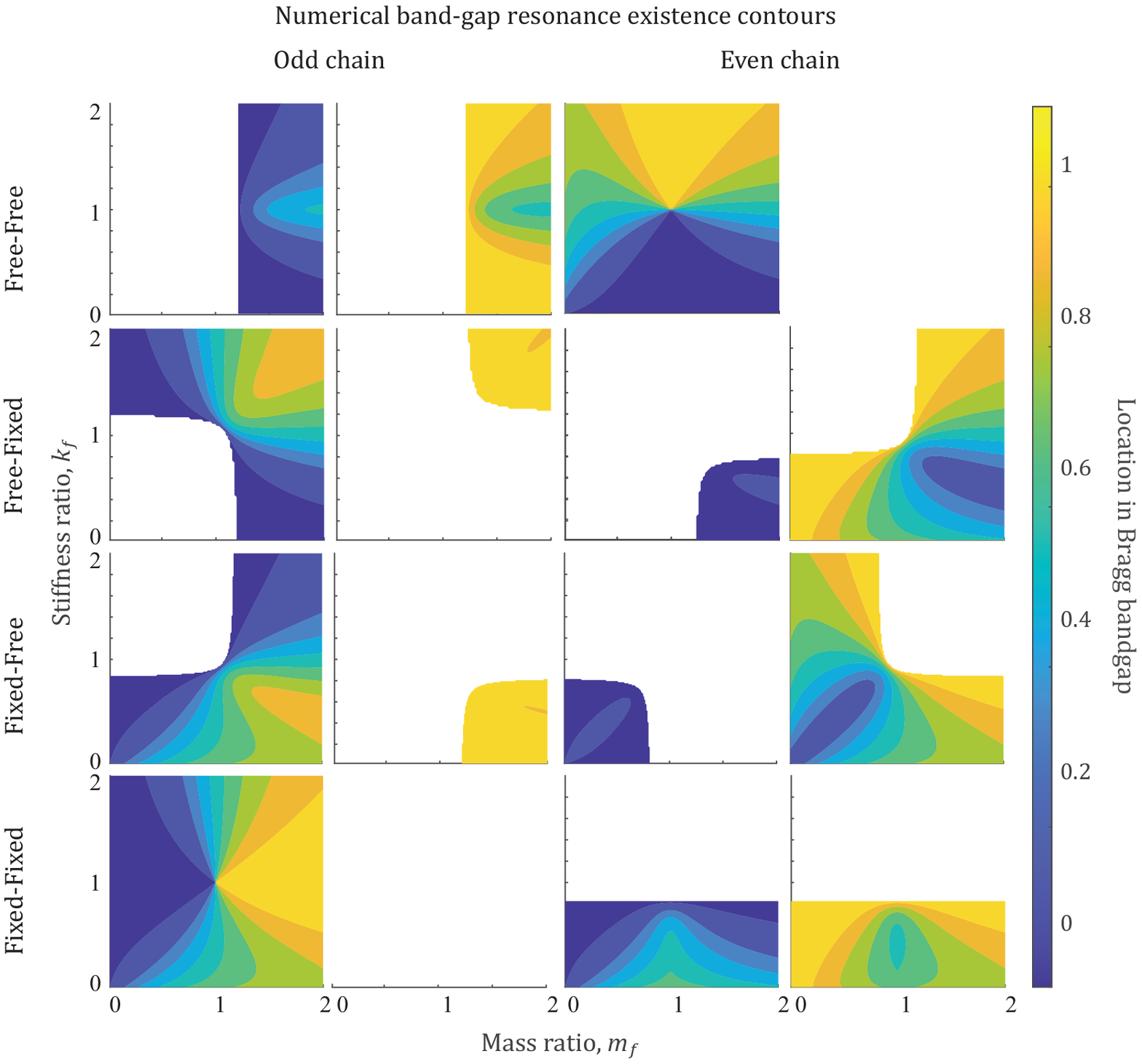}
             \caption{Numerical contours for the relative location of band-gap resonances for 5-unit-cell odd and even chains with free-free, free-fixed, fixed-free and fixed-fixed boundaries as a function of the diatomic unit cell mass and stiffness ratios $m_f$ and $k_f$. Blue contours indicate resonances close to the acoustic branch (lower end of the band gap), while yellow contours indicate resonances near the optical branch (upper end of the band gap).}
        \label{fig:NumRegions}
\end{figure*}

Following this approach, the conditions of existence of band-gap resonances in odd and even chains with free or fixed boundaries are listed in \autoref{Tab:ExistenceConditions}. 
A graphical representation is also provided in \autoref{fig:AnalRegions} for these conditions as functions of $k_f$ and $m_f$ for an example in which the number of the diatomic-chain unit cells is $n = 5$. In this figure, the acoustic-branch crossing curves are indicated by the solid lines and the optical-branch crossing curves are indicated by the dashed lines.
The analytically derived band-gap existence conditions in \autoref{Tab:ExistenceConditions} and \autoref{fig:AnalRegions} are numerically verified in \autoref{fig:NumRegions}. We note that the derived band-gap resonances' existence conditions for free-free chains match Wallis's findings for the case of a diatomic chain with a unit cell that has two springs with equal constants \cite{wallis1957effect}. 
Moreover, experimental and numerical studies on band-gap resonance frequencies conducted by A.C. Hladky-Hennion et al. \cite{hladky2005localized,hladky2007experimental} obtained the same existence conditions for free-free diatomic chains composed of welded or glued beads, when examined in the low-frequency regime. 
These beads exhibit no internal deformation as they move like rigid bodies relative to each other in the low-frequency regime, making a discrete mass-spring model appropriate. The discrete model's resonance frequencies are within a maximum of $4.5\%$ error from their experimentally observed counterparts. 
These observations point to the fact that the presented formulation and results apply to one-dimensional periodic granular materials in the near-linear regime$-$hence demonstrating direct applicability to the analysis and design of this intriguing class of phononic media.\\
\begin{table}[hb!]
\centering
\begin{minipage}{40em}
    \caption{Existence conditions for odd and even diatomic chains with fixed or free boundary conditions at each end}
    \centering
      \setlength{\tabcolsep}{4pt}
  \begin{tabular}{c|lc}
   \textbf{Parity} & \textbf{BCs} & \textbf{Region equation}  \\ \hline \hline
   \multirow{4}{*}{Odd} & Free-free  & $m_f > \frac{n+1}{n} $\\ 
   & Free-Fixed & $k_f > \frac{(1+n)(2n(m_f-1)+m_f-3)}{(1+2n)(n(m_f-1)-1)}$\\
   & Fixed-Free & $k_f < \frac{(1+2n)(n(m_f-1)-1)}{(1+n)(m_f(1+2n)-2n-3)}$  \\ 
   & Fixed-Fixed & $\Omega^2 = 1 + k_f$\\
   \hline
   \multirow{4}{*}{Even} & Free-free  & $\Omega^2 = \frac{k_f (1 + m_f)}{m_f}$\\ 
   & Free-Fixed & $k_f < \frac{n(2n(m_f-1)-m_f-1)}{(1+2n)(n(m_f-1)-1)}$\\
   & Fixed-Free & $k_f < \frac{n (1+m_f - 2n(1-m_f))}{(1+2n)(m_f(1+n)-n)}$ \\ 
   & Fixed-Fixed & $k_f < \frac{n}{n+1}$ 
    \label{Tab:ExistenceConditions}
    \end{tabular}   
\end{minipage}
\end{table} 
\indent A particularly useful feature of this approach is that asymptotic limits and critical points in the hyper-surface for different problem parameters, e.g., $m_f$ or $k_f$, are easily provided for application-specific design processes.
These points are important in determining design preliminary criteria if we want to target regions where there might be two, one, or no band-gap resonance frequencies for specific unit-cell number $n$ and boundary conditions. 
For instance, in the case of a fixed-free even chain, the acoustic-branch border curve intersects $m_f = 0$ at $k_f = {(2n-1)}/{(2n+1)}$ and intersects $k_f = 0$ at $m_f = {(2n-1)}/{(2n+1)}$. 
Moreover, the optical-branch curve is asymptotic to $k_f = {n}/{(n+1)}$ as $m_f \to \infty$ and to $m_f = {n}/{(n+1)}$ as $k_f \to \infty$. 
These limits and points are obtained for all cases and listed in \autoref{Tab:Asymptotes}.

Finally, inspecting the existence regions in \autoref{fig:AnalRegions}, as well as the critical points and asymptotes in \autoref{Tab:Asymptotes}, some observations can be made:
\begin{itemize}
\item In general, band-gap resonances exist when the lighter mass$-$or the stiffer spring$-$is at the free boundary, or when the heavier mass$-$or the softer spring$-$is at the fixed boundary. 

This can be seen by examining the existence regions for free-free odd chains and fixed-fixed even chains.
Another example can also be seen upon comparing the existence regions for free-fixed and fixed-free even chains [see \autoref{fig:introX}]. Both chains are connected to the rigid wall by a spring constant of $k_1$, which excludes the determining effect of the spring constant. 
Thus, band-gap resonances exist when the heavier mass is connected to the rigid wall. 
This condition is satisfied when $m_f>1$ for the free-fixed even chain and $m_f<1$ for the fixed-free even chain. 
Therefore, the existence regions for these two cases mirror each other about the $m_f = 1$ axis. 
The same reasoning explains why the existence regions for odd chains with fixed-free and free-fixed boundary conditions mirror each other about the $k_f = 1$ axis.

\item The presence of the lighter mass at the boundary is the most important determining factor for the existence of band-gap resonances in the case of free boundaries, while the presence of the stiffer spring at the rigid-wall boundary is the most important determining factor in the case of fixed boundaries. 

For instance, the presence of the lighter masses at the boundaries in odd free-free chains exclusively determines the presence of band-gap resonances for all possible stiffness ratios $k_f$. The same observation is made concerning even fixed-fixed chains; the presence of the stiffer springs at the boundaries exclusively determines the presence of the band-gap resonances for all possible mass ratios $m_f$. Notice how the fixed-fixed odd-chain existence regions are a $90^\circ$-clockwise rotated version of the free-free even-chain ones, indicating this switched role between the masses and springs in the free-boundary and fixed-boundary cases, respectively.  
\end{itemize}

\begin{table}[htbp!]
\centering
\begin{minipage}{40em}
    \caption{Limits and asymptotes for existence conditions of diatomic chains with free or fixed boundaries at each end (see  \autoref{fig:AnalRegions})}
    \centering
      \setlength{\tabcolsep}{3pt}
  \begin{tabular}{c|c|cccc}
  \multirow{2}{*} {\textbf{Parity}} & \multirow{2}{*} {\textbf{BCs}} & \multicolumn{2}{c}{$m_f$} & \multicolumn{2}{c}{$k_f$}  \\

& & $k_f = 0$ &  $k_f \to \infty$ & $m_f = 0$ &  $m_f \to \infty$\\
   
    \hline \hline
   
   \multirow{4}{*}{Odd} & Free-Free  & $\frac{n+1}{n}$ & $\frac{n+1}{n}$ & - & -\\ 
   & Free-Fixed & $\frac{2n+3}{2n+1}$ & $\frac{n+1}{n}$ & $\frac{2n+3}{2n+1}$ & $\frac{n+1}{n}$\\
   & Fixed-Free & $\frac{n+1}{n}$ & $\frac{2n+3}{2n+1}$ & $\frac{2n+1}{2n+3}$ & $\frac{n}{n+1}$  \\ 
   & Fixed-Fixed & - & - &  - & -\\
   \hline
   \multirow{4}{*}{Even} & Free-Free  & - & - &  - & - \\ 
   & Free-Fixed & $\frac{2n+1}{2n-1}$ & $\frac{n+1}{n}$ & $\frac{n}{n+1}$ & $\frac{2n-1}{2n+1}$ \\
   & Fixed-Free & $\frac{2n-1}{2n+1}$ & $\frac{n}{n+1}$ & $\frac{2n-1}{2n+1}$ & $\frac{n}{n+1}$ \\ 
   & Fixed-Fixed & - & - & $\frac{n}{n+1}$ & $\frac{n}{n+1}$
    \label{Tab:Asymptotes}
    \end{tabular}   
\end{minipage}
\end{table}

\section{Example of altered structure: Finite chain with even parity and an arbitrary mass at one end}
\label{sec:differentBoundaries}

\begin{figure*}[htbp!]
\centering
\centering
        \includegraphics{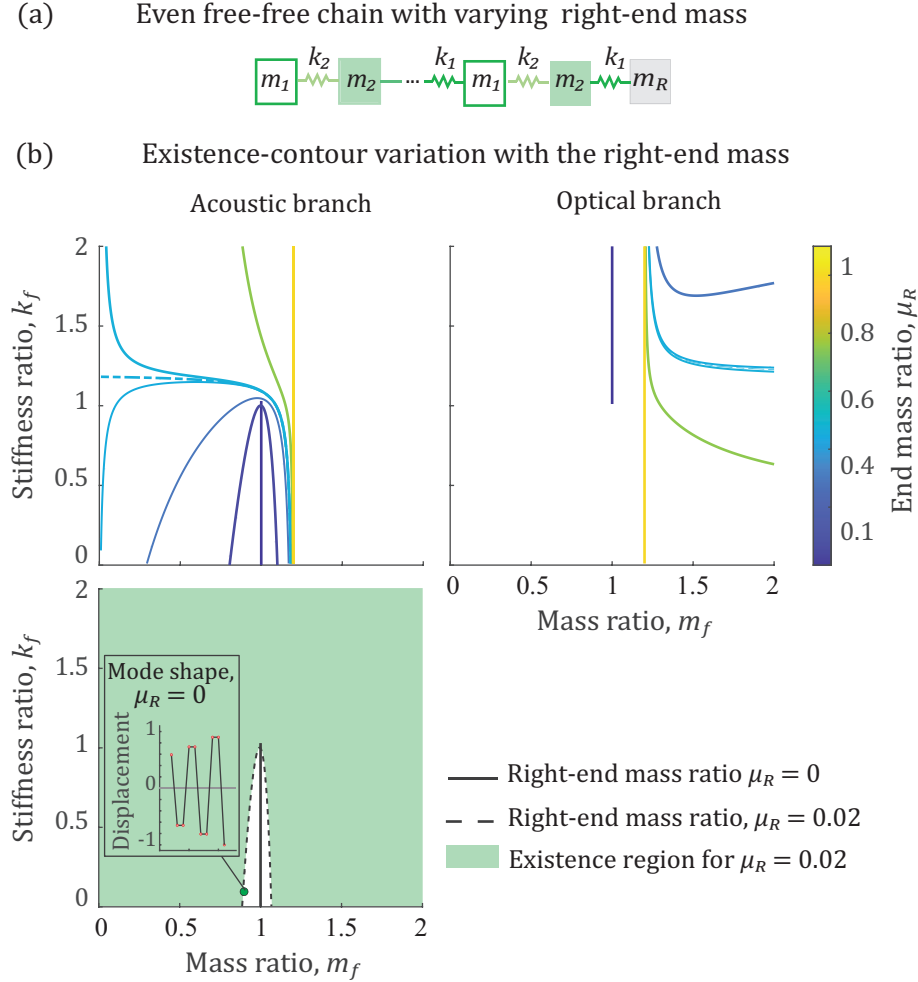}
             \caption{\textbf{(a)} A free-free even chain with a variable right-end mass $m_\text{R}$. \textbf{(b)} The existence regions for a case in which $\mu_\text{R} = 0.02$. One point ($m_f = 0.9$, $k_f = 0.1$) is highlighted which lies in the band-gap region when $\mu_\text{R}=0$ (mode shape is drawn in inset; existence regions are shown in \autoref{fig:AnalRegions}). This point becomes a band-gap edge point when $\mu_\text{R} = 0.02$. \textbf{(c)} Border curves of band-gap resonances existence regions for a 5-unit cell numerical example with different values of the right-end mass ratio $\mu_\text{R} = {m_\text{R}}/{m_1}$, }
        \label{fig:VariableMass}
\end{figure*}

The second special case we consider is that of an even chain with a boundary mass attached to the right end [\autoref{fig:VariableMass}(a)]. The characteristic equation for this system can be obtained by setting the left-boundary-mass parameter $\mu_\text{L}$ to $0$ in the even-chain case (a) characteristic-equation coefficients [\autoref{Tab:CharCoeffValues}]. The resulting coefficients of the characteristic equation in Eq. \eqref{eq:charGen} become
\begin{subequations}
\begin{gather}
A = \csc \theta \\
c_2 = k_f^2\\
c_1 = - k_f m_f \Omega^4 +  k_f \Omega^2 \left( 1+ k_f + m_f + k_f m_f - \mu_\text{R} \right) - 2k_f^2 \\
c_0 = (k_f - m_f \Omega^2) (k_f - (1-\mu_\text{R})\Omega^2).
\end{gather}
\label{eq:CaseIICharEqn}
\end{subequations}
Taking the limit of the characteristic equation as $\theta \to \pi$, we find that
\begin{gather}
k_f^2 (1+n) (\Omega^2 - 4 + m_f \Omega^2) - k_f \Omega^2 (m_f (n-1 (\Omega^2-2)+\Omega^2) \nonumber\\
+ (1+2n)(\mu_\text{R}-1)) + m_f n \Omega^2 (\mu_\text{R} - 1) = 0.
\label{eq:CaseIIIntEq}
\end{gather}
Combining Eq. \eqref{eq:CaseIIIntEq} with the IBZ-edge frequencies in Eq. \eqref{eq:IBZedgeSquared}, the hyper-surface equation can be computed (too lengthy to be included here). In \autoref{fig:VariableMass}(b), it is shown that even perturbing the right-boundary mass by $2\%$ of the value of $m_1$ would result in converting band-gap modes [see mode shape in \autoref{fig:VariableMass}(b)] to pass-band modes. 

Considering several values for the right-end mass ratio $\mu_\text{R}$, the corresponding border curves are plotted in \autoref{fig:VariableMass}(c). 
At $\mu_\text{R} = 0$, the problem is essentially that of a free-free even-chain case and the acoustic- and optical-branch crossing lines are the same as in \autoref{fig:AnalRegions}.  In the free-free even-chain case, there is only one band-gap resonance that exists, which is close to either the acoustic or the optical branches, depending on the values of $m_f$ and $k_f$. An explicit expression exists for this case and is written in Eq. \eqref{eq:EvenFF_Stop}. 
On the other hand, at $\mu_\text{R} = 1$, the problem is that of a free-free odd-chain case; thus two band-gap resonance frequencies cross from the the acoustic and optical branches when $m_f > {(n+1)}/{n}$. 
It is noted that the acoustic and optical branches' border curves are separated here into two figures for clarity, as opposed to being represented in the same plot as in \autoref{fig:AnalRegions}. 

Varying the right-end mass ratio $\mu_\text{R}$ between $0$ and $1$ highlights the even-to-odd chain transition from the point of view of the band-gap resonances' existence conditions in free-free diatomic chains. Considering the acoustic-branch border curves, their right end always has an intersection point with $k_f = 0$. This intersection point moves gradually from 1 (even-chain case) to ${(n+1)}/{n}$ (odd-chain case) as $\mu_\text{R}$ increases. 
Meanwhile, the left ends of the border curves intersect with $k_f = 0$ as $\mu_\text{R}$ increases, up to a critical transition point at $\mu_\text{R} = {n}/{(2n+1)}$. Note that $m_f$ never reaches zero in these curves except when $\mu_\text{R}$ reaches its critical value. The $(m_f, k_f)$ coordinates of this critical value can be obtained by inspecting the trends of the change in the border curves and computing  
\begin{equation}
\lim_{m_f \to 0} k_f(m_f,n.\mu_\text{R})  = \frac{2n+3}{2n+1}.
\label{eq:Critkf}
\end{equation} 
This marks a clear transition point in the trend of the border curves as can be seen in \autoref{fig:VariableMass}(c). 
This continues to gradually change as $\mu_\text{R}$ approaches 1, at which point no band-gap resonances exist at $m_f<{(n+1)}/{n}$ for any value of $k_f$. 
On the other hand, the optical branch exhibits a sudden transition that occurs with perturbing $\mu_\text{R}$. It is noted that as $\mu_\text{R}$ is perturbed, two band-gap resonances can now exist, instead of only one. The border curves are asymptotic on one end to $m_f = {(n+1)}/{n}$ as $k_f \to \infty$. As the value of $\mu_\text{R}$ increases, the other end of the border curves moves down as they gradually transform into the border line for the free-free odd chain.  
Therefore, critical points for the transition between border curves at different chain parity can be identified. Furthermore, the same procedure can be performed to investigate the effect of boundary conditions on band-gap resonances' existence conditions as the end-masses are varied from zero to infinity to emulate the free-to-fixed boundary-condition transition. This facilitates the control of these band-gap resonances, and opens up the door for multiple applications that would particularly benefit from precise control of their properties, e.g., flow stabilization using phononic subsurfaces where the isolation of the resonances enhances control robustness \cite{hussein2015flow}, and mass sensing where the degree of isolation affects probe sensitivity \cite{bonhomme2019love} [see the example in \autoref{fig:VariableMass}(b)].

\section{Conclusions}
\label{sec:Conclusions}

In this paper, closed-form existence conditions for band-gap resonance frequencies in discrete diatomic chains are presented. Our approach is applicable to (1) general boundary conditions, (2) odd or even chain parity, and cases where a unique mass and/or spring component is placed at the ends of the truncated chain. The formulation to obtain these conditions may be easily extended to periodic chains with different unit-cell compositions, as long as the system satisfies a $k$-Toeplitz matrix. The existence conditions are obtained by exploiting the presence of the wave number as a variable in the finite system's characteristic equation, and taking the characteristic equation's limit as the wave number approaches the IRB edge.
A special case is shown where the band-gap resonances existence conditions for chains with free-fixed, fixed-free, and fixed-fixed boundary conditions is presented for the first time. 
The results are numerically verified, and are also found to match computational and experimental findings in the literature for free-free chains made of glued or welded beads. 
Furthermore, the systematic nature of this approach enables us to navigate even more complex cases and boundary conditions.
For example, a special case is shown where a variable right-end mass is attached to an even free-free chain. 
Gradual transition of the existence conditions is observed, starting from the case of an even free-free chain and ending with that of an odd free-free chain. Critical transition points and limits are identified as functions of the number of unit cells in the chain and the mass and stiffness ratios in the unit cell. 
A numerical example is provided that shows that band-gap resonances disappear from an even free-free chain when the right-end mass is perturbed by only $2\%$ of the values of the first-mass in the diatomic unit cell. 
This ability to systematically control the existence of gap resonances is beneficial for a variety of applications. It could potentially be used also for active control where, for example, a computer program would spontaneously determine the band-gap resonances' existence conditions in real-time during the operation of a finite-sized phononic-crystal-based device.

\begin{acknowledgments}
This research was partially supported by the Advanced Research Projects Agency-Energy (ARPA-E) grant number DE-AR0001056.
\end{acknowledgments}

\bibliographystyle{elsarticle-num-names}
\bibliography{litrev}




\appendix

\end{document}